\documentclass[review]{elsarticle}
\usepackage{hyperref}
\usepackage{amsmath}
\usepackage{longtable}
\usepackage{epstopdf}  

\begin{document}

\begin{frontmatter}

\title{X-Ray Spectroscopy: An Experimental Technique to Measure Charge State Distribution Right at the Ion-Solid Interaction}


\author[]{Prashant Sharma} \corref{cor1}

\author[]{Tapan Nandi}

\address{Inter University Accelerator Centre, Aruna Asaf Ali Marg, New Delhi - 110067, INDIA}


\begin{abstract}
Charge state distributions of $^{56}$Fe and $^{58}$Ni projectile ions passing through thin carbon foils have been studied in the energy range of 1.44 - 2.69 MeV/u using a novel method from the x-ray spectroscopy technique.  Interestingly the charge state distribution in the bulk show Lorentzian behavior instead of usual Gaussian distribution. Further, different parameters of charge state distribution like mean charge state, distribution width and asymmetric parameter are determined and compared with the  empirical calculations and ETACHA predictions. It is found that the x-ray measurement technique is appropriate to determine the mean charge state right at the interaction zone or in the bulk. Interestingly, empirical formalism predicts much lower projectile mean charge states compare to x-ray measurements which clearly indicate multi-electron capture from the target surface. The ETACHA predictions and experimental results are found to be comparable for energies $\geq$ 2 MeV/u.
\end{abstract}

\begin{keyword}
\texttt{ X-ray spectroscopy \sep Electron capture and loss process \sep Charge state distribution \sep Ion-solid collision}

\end{keyword}

\end{frontmatter}


\section{Introduction}

When an ion beam passes through the target a considerable amount of fluctuation takes place in charge state of the ion due to different atomic phenomena e.g. electron capture and loss processes, inner-shell ionization etc. occurring both in the bulk and at the target surface. In general, final charge state of the ion depends not only on the parameters associated with incident ion (initial ion velocity, charge and atomic number) but also on target characteristics (atomic number and density). Even though a monochromatic ion beam with a fixed charge state is passed through the medium, several charge states emerge out of the target \cite{shima92}. However, after a large number of collisions, an equilibrium in charge state distribution (CSD) is established, where certain balance in electron capture and loss processes are attained. The study of equilibrium conditions where charge state fractions (CSF) as well as mean charge state ($q_{m}$) reach to a certain stability are very crucial and have plenty of applications in various fields including atomic physics \cite{baud80}, nuclear physics \cite{sko8}, astrophysics \cite{liu14}, biophysics \cite{black10}, energy loss experiments \cite{Meyer}, accelerator designs \cite{Niello04}, detectors \cite{Taniike}. Many experimental and theoretical groups have worked in these fields of research since 1950's \cite{Stier} and therefore many reviews or collection of data can be found in the literature based on electromagnetic methods and empirical formalisms (e.g. Allison \cite{Allison}, Betz \cite{Betz}, Wittkower \cite{Wittkower}, and Shima \cite{shima92,shima86}).

Worth to mention that there are many experimental techniques like CRBS \cite{Sa'adeh}, recoil separator \cite{Leino,Khuyagbaatar}, TOF \cite{Dickel2015}, electromagnetic methods \cite{Academy1982} etc. which can be used to separate adjacent charge states and to study the charge state distribution of the respective atomic systems. The main demerit of these techniques is that they account for the total charge of the ion in the detectors placed a few meters away from the target. This implies that these techniques give a measure of electron-capture and -loss processes in bulk as well as at surface of the foil and cannot segregate the charge changing phenomenon occurring only in the bulk or of the surface. Therefore, these techniques are not appropriate to measure the charge state distribution right at the ion-solid interaction. This difficulty can be avoided by using charge less observables in the experiments to measure the charge states right at the ion-solid interaction. Interestingly, in the past several measurements have been carried out using x-ray spectroscopy to study various plasmas \cite{Santos,Ullmann}. Nevertheless, the measurements with x-ray photon detection have not yet been employed to study the CSD and other relevant parameters $q_{m}$, distribution width and asymmetric parameter of the projectile ions during passage from any solid/gas targets.
With this motivation, we confine the work to study the CSD  and its parameters right at the ion-solid interaction zone using the x-ray spectroscopy technique.

\section{Experiments}

The experiments were performed with the energetic ion beams of $^{56}$Fe and $^{58}$Ni using 15 UD Pelletron \cite{Kanjilal}  accelerator at IUAC, New Delhi. Well-collimated ion beam in the energy range of 1.44 - 2.69 MeV/u were bombarded on 80 $\mu$g/cm$^2$ ($\approx$ 113 $\mu$g/cm$^2$ at 45$^{\circ}$) thick amorphous carbon target foils to produce the equilibrium charge state distribution. The target was placed at 45$^{\circ}$ to the beam axis so that the x-ray spectra could be measured right from the ion-solid interaction zone. The x-rays were detected in a Low Energy Germanium Detector (GUL0035, Canberra Inc., with 25 $\mu$m thick Be entrance window, resolution 150 eV at 5.9 keV) placed at 90$^{\circ}$ to the beam axis to avoid the Doppler shift. The x-ray produced in the ion-solid interactions were passed through two collimators of 3 mm diameter kept at 55 cm apart whereas the first collimator was placed at 10 cm away from the target. This configuration ensured that the x-rays were coming from a tiny section ($\pm$3-4 mm) of the ion-solid interaction zone. In the time scale, the x-ray detector could observe only atomic transitions of very short life-time (few tens of psec) with respect to the centre of the interaction zone. Hence, the x-ray spectroscopy technique could be considered as a measurement  right at the ion-solid interaction zone compared to the electromagnetic measurements taking place away from the interaction zone or at t $\approx$ a few $\mu$sec for MeV ions. The x-ray detector was placed outside the chamber at 65 cm away from the target separating a thin mylar window of 6 $\mu$m at the interface of detector and chamber. The beam was dumped in a Faraday cup. Two silicon surface barrier detectors were used at $\pm$10$^{\circ}$ to monitor the beam direction. The x-ray spectra observed for all the beam energies are shown in Fig. 1. Calibrations were done for the x-ray detectors using $^{60}$Co and $^{241}$Am standard radioactive sources. The resolution was found to be about 200 eV at 6.41 keV with the experimental conditions in the beam hall. Further, the calibration was internally verified through Fe $K_{\alpha}$ and $K_{\beta}$ peaks due to beam halo hitting the carbon foil holder made up of stainless steel in the case of $^{58}$Ni projectile ions. However, in the case of $^{56}$Fe-beam experiment, beam halo was minimized by passing the beam through a blank target frame so that its presence did not affect much the peak structure originated from the projectile ions. Vacuum chamber was maintained at a pressure around 1$\times$10$^{-6}$ Torr.	
 \begin{figure*}[!ht]

\includegraphics[scale=.8]{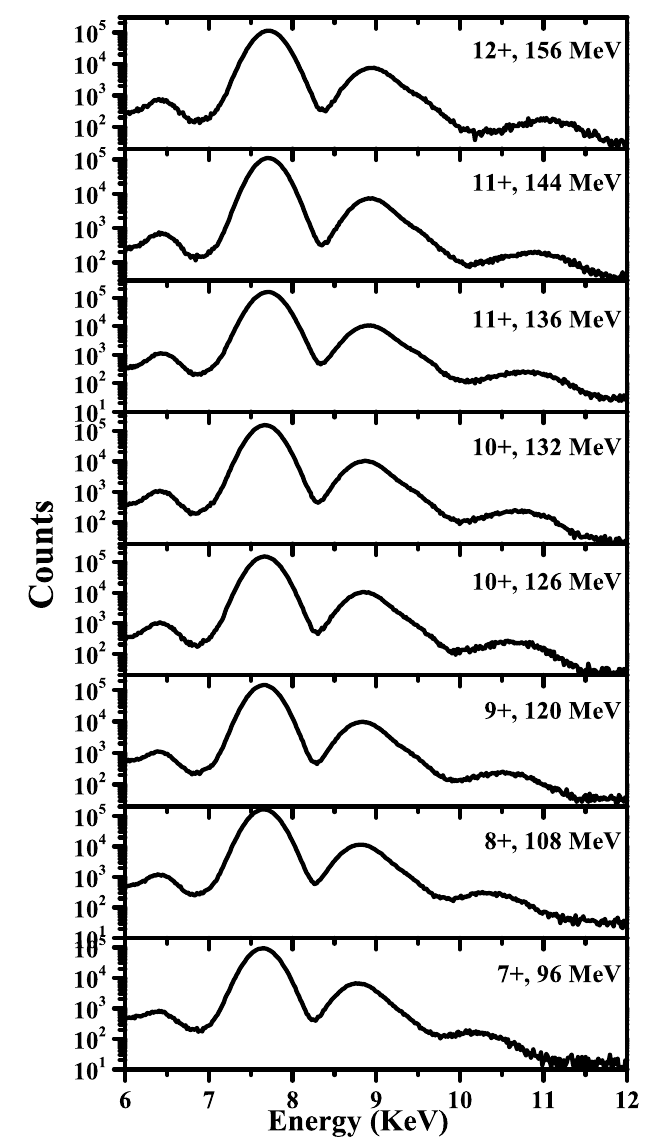}
\includegraphics[scale=.8]{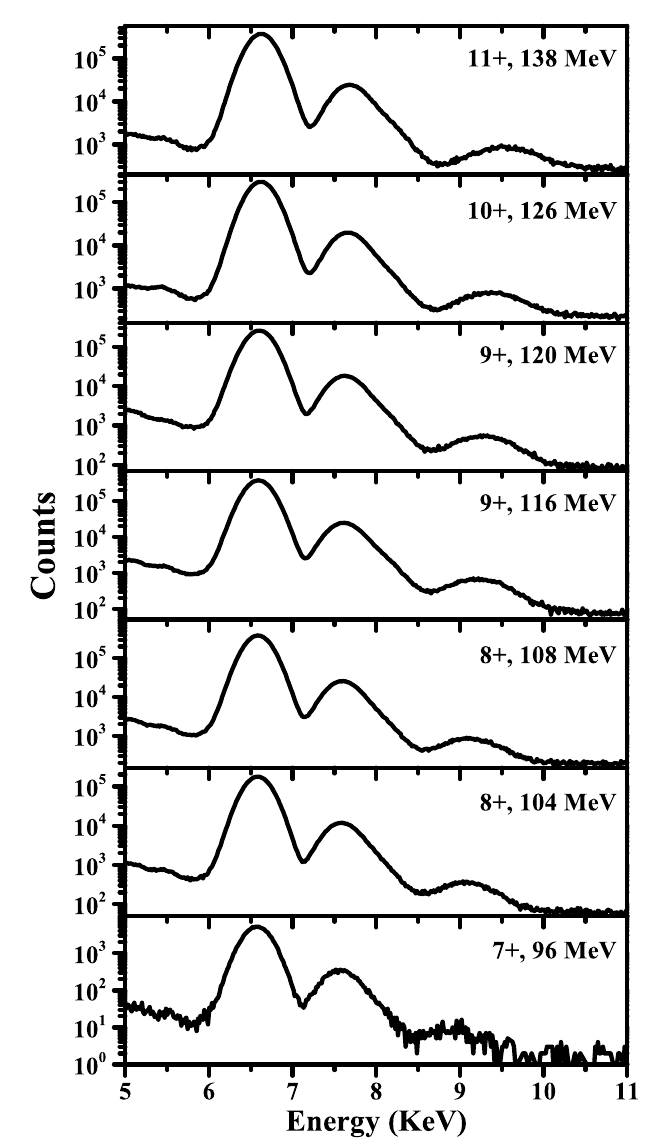}

\caption{\label{fig1} X-ray spectra for (a) $^{58}$Ni beam and (b) $^{56}$Fe beam on 80 $\mu$g/cm$^{2}$ C-foil at different beam energies and initial charge states
}
\end{figure*}
\section{Data Analysis and Results}

In this work we are intended to determine the charge state distribution of projectile ions right at the ion-solid interaction along with $q_{m}$ and other relevant parameters from the measured x-ray spectra. Accordingly we develop a novel method to extract required information from the x-ray spectra observed. Worth mentioning here that the parameters like charge state fraction, mean charge state etc. obtained at different energies for the particular ion can be compared without normalizing the x-ray spectra. Hence, normalization of x-ray spectra is not carried out in this work like any other electromagnetic methods coupled with position sensitive detectors \cite{Stöhlker}.
\begin{figure}[!ht]

\includegraphics[scale=.42]{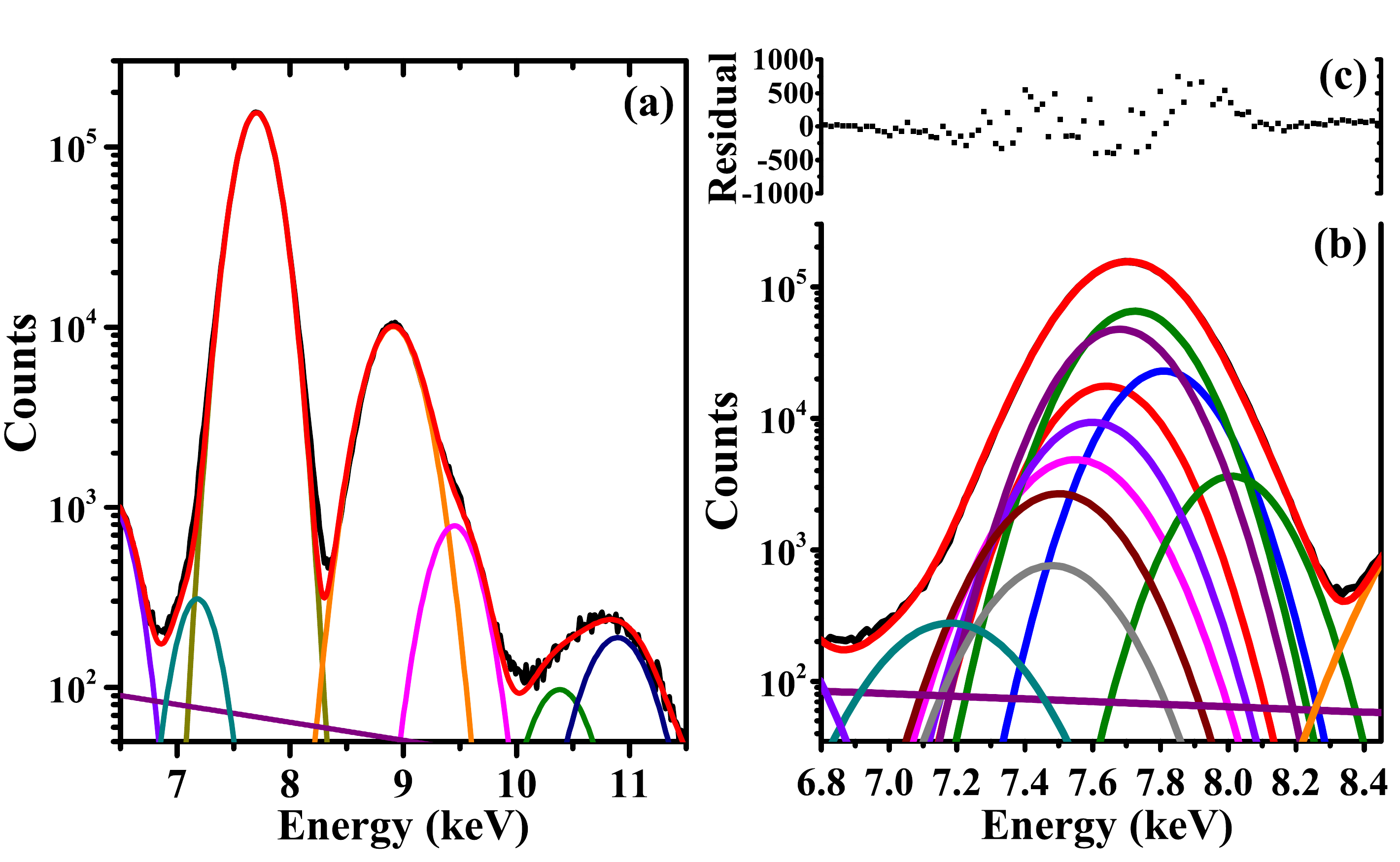}

\caption{\label{fig2} X-ray spectrum of $^{58}$Ni on C at 136 MeV (a) Fitting shows only broad features of the spectrum (b) Projectile x-ray peak is fitted into nine Gaussian functions corresponding to x-ray lines appearing from H-like to F-like $^{58}$Ni with a exponential function representing the background (c) The residuals of fitting (b) is shown }

\end{figure}
It is clear from the spectrum of 136 MeV $^{58}$Ni on C that it contains mainly three structures as shown in Fig. \ref{fig1} and \ref{fig2}a. In our earlier work the first peak (7 - 8.4 keV) is recognized to have originated from the projectile ion x-ray, whereas second and third peak (8.4 - 11.5 keV) belong to the projectile-like fragment ions emanating from the nuclear reactions, respectively \cite{Nandi09,Ahmad05}. It is worth mentioning here that we are only considering the charge changing phenomena in elastic events or in the projectile ions, thus the second and third structures are of no relevance in this work; hence they will not be brought in the further discussion. Another point is to be cleared at this stage that when the fast projectile ions incident on target atom, the collisions create vacancies in the different shells of both target atoms and projectile ions, which cause shift in the characteristic x-ray line energies \cite{Kumar,Schnopper}.
\begin{figure*}[!ht]

\includegraphics[scale=.315]{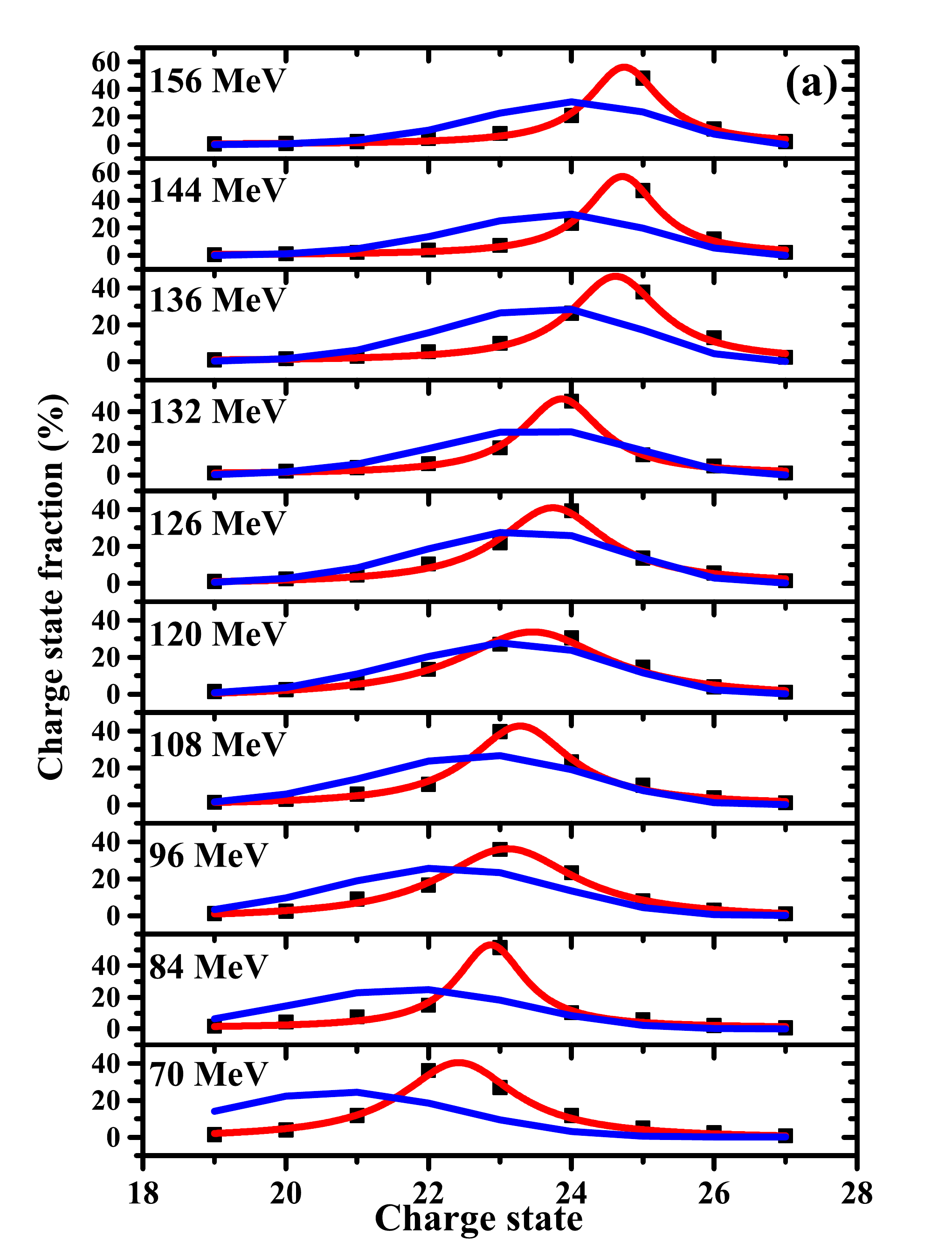}
\includegraphics[scale=.79]{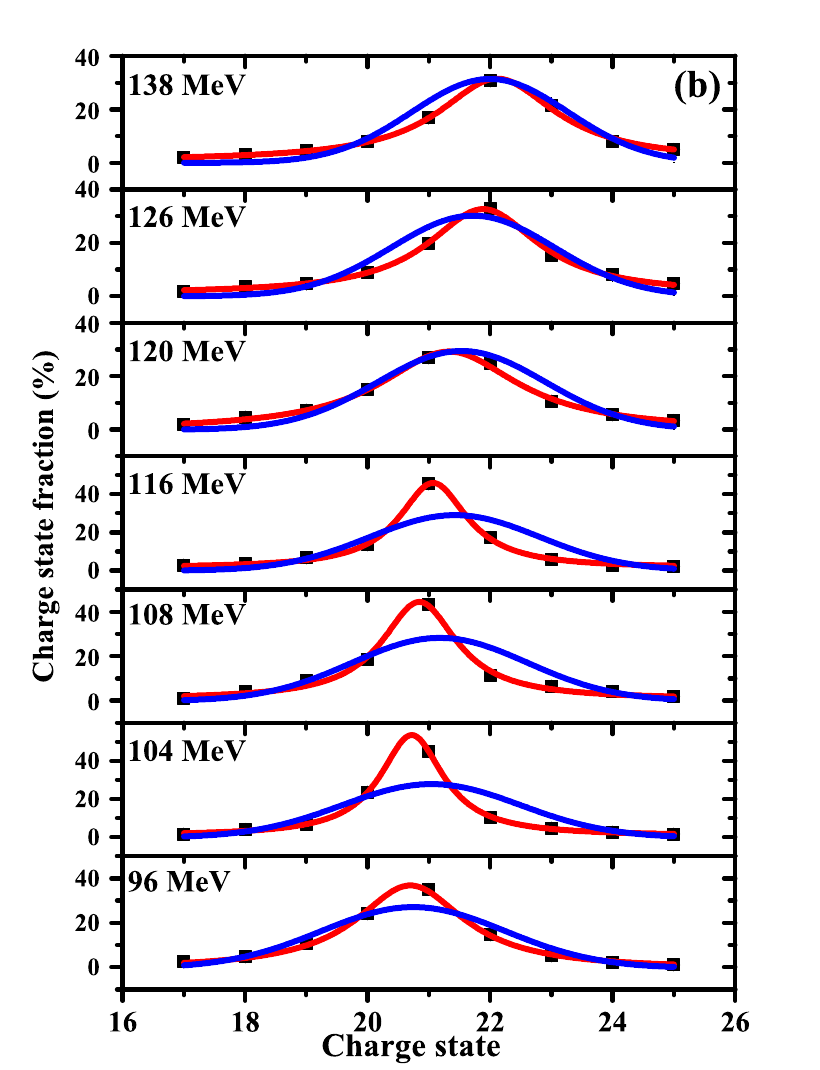}
\caption{\label{fig3} Comparison between experimental and ETACHA charge state distribution for (a) $^{58}$Ni on C (b) $^{56}$Fe on C for various beam energies. Figure is showing Lorentzian fit (red) to experimental data and Gaussian fitting (blue)  to ETACHA data \cite{Rozet}). Due to small values, errors are embedded in symbol itself.
}
\end{figure*}
 The static target atoms are completely filled before the collisions, so the shift in K-shell x-ray energies depends on vacancies created in K-, L-, M- shells etc. during ion-solid interactions. In contrast, in the case of projectile ions the outer electrons get stripped completely up to a certain level during ion-solid collision (depending upon the collision energy) \cite{Schnopper} and the line energy shift solely depends on K- and L-shell vacancy only due to absence of  electrons in the M-, N- and further shells \cite{Kumar}. The increased ionic charge of the projectile ions shift the centroid towards higher energy side with the increase in beam energies. Such a scenario is clearly depicted in Fig. \ref{fig1}. 

 Important to note that life-time of a single K-vacancy is of the order of few fsec. Whereas in the intermediate energy regime (this work), projectile ion takes a few tens of fsec before leaving the target thus encountering multiple collisions. It causes a complex chain of collision events resulting in a composition of many charge states under the projectile x-ray peak. Though poor detector resolution constrains us to resolving the individual x-ray lines but the well-defined centroid gives a measure of the mean charge state of the projectile ions. To find the mean charge state and charge state distribution; projectile x-ray peak has been fitted with various Gaussian peaks ranging centroid energy for H-like to F-like ions, as shown in Fig. 2(b). It should be noted here that $K_{\alpha}$ x-ray energies for Ni are available only for H-like to Li-like Ni \cite{Ralchenko}. The rest of the energies for Be-like to F-like Ni are scaled from corresponding Fe data \cite{Ralchenko}. 

In this way we have obtained the distribution of CSFs ($F_{q}$) due to K-shell vacancy production directly from the measured x-ray line intensities as follows
\begin{equation} \label{Eqa}
F_q = \frac{I_q\omega_q}{\sum_q I_q \omega_q}
\end{equation}

\noindent here, $I_q$ and $\omega_q$ corresponds to intensity and fluorescence yield of the projectile ion in the final electronic configuration, respectively.
Noteworthy that, for the values of fluorescence yield, one needs to know the vacancy distribution in the initial configuration state, corresponding sets of the transitions schemes and the transition rates during heavy ion collisions. This can be easily estimated for target atoms. However, in the case of projectile ions (this case) neither is very well known nor is an easy problem to solve. Recently, Dr. M.F. Haso\u{g}ulu has theoretically calculated LSJ-state dependent fluorescence yields in the case of Li-like to F-like ions in the range of 3$\leq$Z$\leq$30 atomic numbers for single K-vacancy cases \cite{hasogulu}.  We have taken the fluorescence yield data for different charge states from their calculations to estimate the fluorescence yield for a particular ionic state by statistical averaging of most probable LSJ states during ion-solid collisions, as discussed in detail elsewhere \cite{Hasoglu2006}. The errors given for the CSF data is a combination of the systematical error (estimated, 5\%) and the statistical error of the measured CSF. This procedure is repeated for all beam energies for both the ion species. All the measured as well as ETACHA predicted charge state distributions at each beam energy have been plotted in Fig. \ref{fig3}. Worthy to mention that the CSDs predicted by ETACHA \cite{Rozet} follow the Gaussian shape, in contrast the experimental values depict a different pattern, which is fitted well with a Lorentzian function for all the cases. Observation of Lorentzian profile for the charge state distribution  is an important characteristic of any plasma \cite{Guerra2013,Foord2004}. This suggests that ion-solid collisions constitute a tenuous high density plasma in the bulk of the solid target, which is described in detail elsewhere \cite{Sharma215}.

\begin{figure*}[!ht]

\includegraphics[scale=.23]{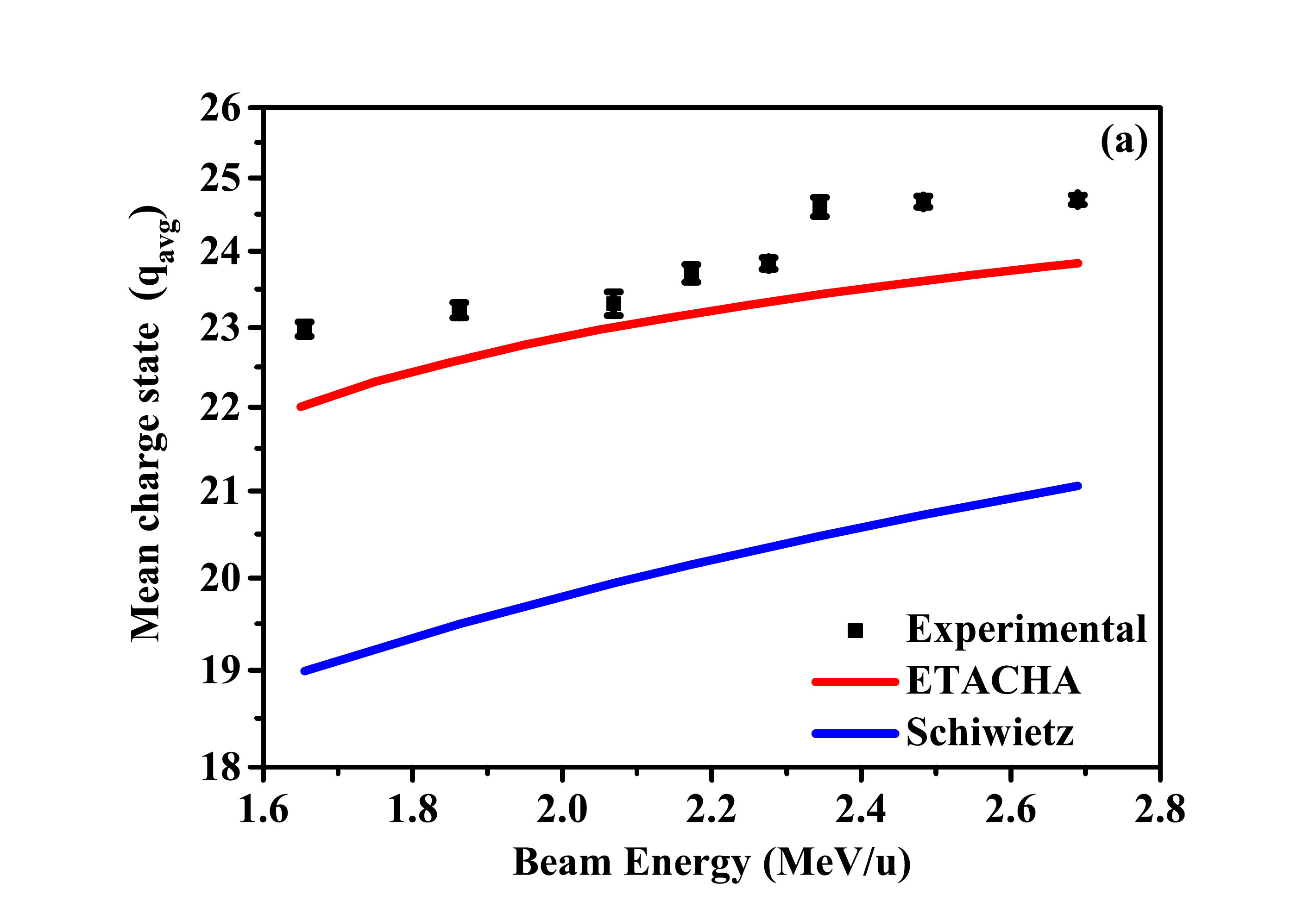}
\includegraphics[scale=.23]{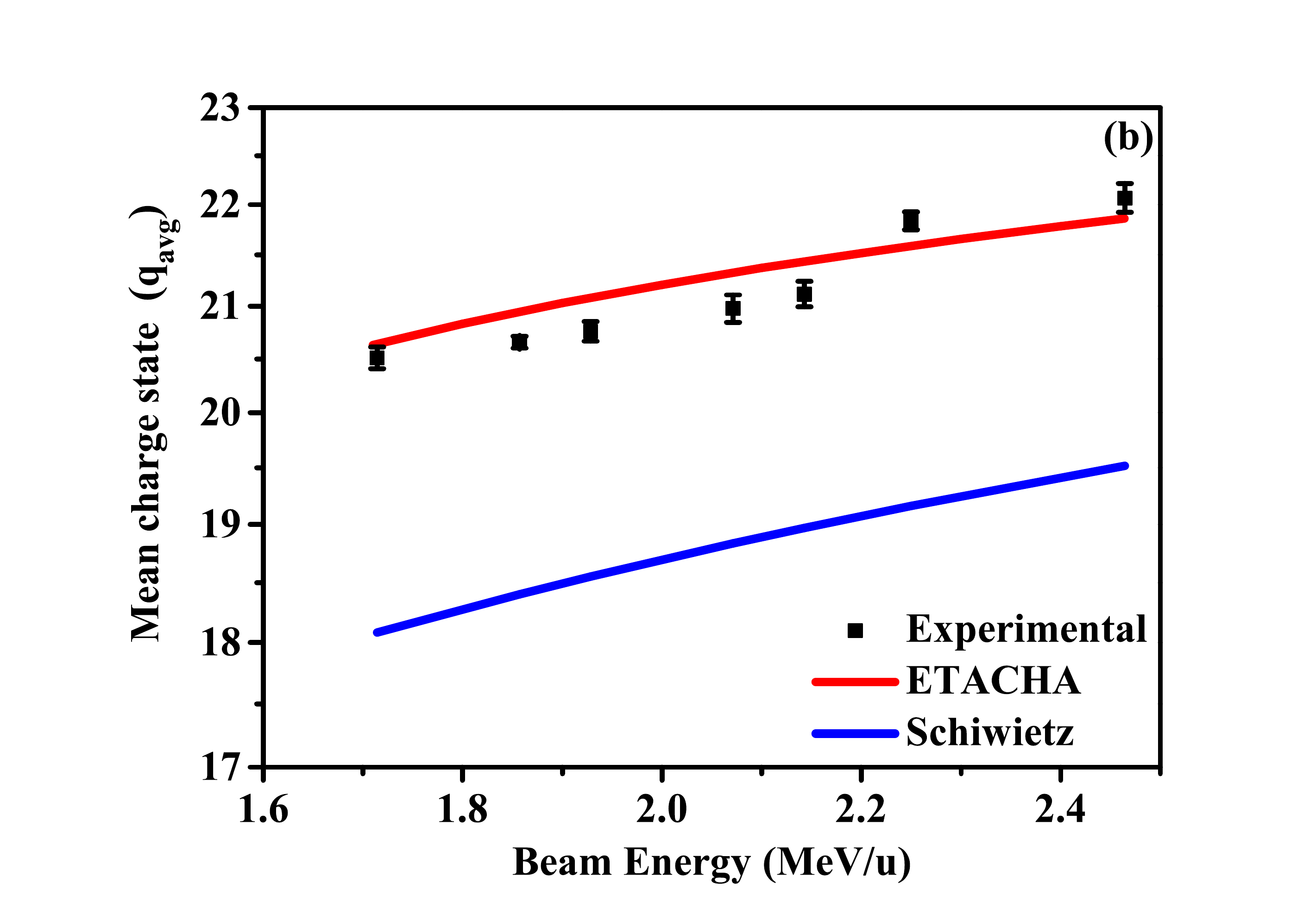}

\caption{\label{figt} Comparison between experimental, ETACHA and Schiwietz mean charge state for (a) $^{58}$Ni on C and (b) $^{56}$Fe on C on 80 $\mu$g/cm$^{2}$ C-foil. Solid lines are to guide eye only. Error bars are smeared within the symbol size.
}
\end{figure*}

Having known the charge state distribution, we step forward to measure the charge state distribution parameters also. Conventionally, the $q_{m}$ is computed using the CSF distribution as follows \cite{Novikov14}
\begin{equation}
q_m = \sum_q q F_q
\end{equation}

\noindent nevertheless, we have adopted a different approach here. Since the centroid of the Lorentzian distribution represents $q_{m}$, we have obtained corresponding $q_{m}$'s from the fitting as shown in Fig. \ref{fig3} and their uncertainties are simply the fitting error of each centroid, $\bigtriangleup q_{m}$. 
Measured $q_{m}$'s are compared to ETACHA calculations \cite{Rozet}\, shown in Fig. \ref{figt}. It is quite clear from the figure that $^{58}$Ni data show some departure from ETACHA predictions \cite{Rozet}, however they agree quite well with each other in case of $^{56}$Fe. Next, for comparison with the available empirical formalisms, we have choose Schiwietz formalism \cite{Schiwietz2001} because of its vast and updated dataset and compared it with experimental results, shown in the same figure. It is found that the empirical formalism shows lower values than both experimental values and ETACHA predictions. 

This is due to the fact that the highly charged ions with the low and intermediate incident energies interact strongly with the target surface. In the earlier work it is reported that projectile ions captures target surface electrons into high Rydberg states to form hollow atoms while entering the target \cite{Briand1990,J.-P.1996,Ninomiya1997}. However, due to the low energy only the interaction with the entrance surface could be studied so far \cite{Brauning2001}. In the intermediate energy region (this work) when the projectile ions leaves the target surface, they can again captures surface electrons into high Rydberg states. However, due to long decay-time (of the order of $\mu$sec-nsec \cite{Gallagher}) these transitions could not be observed using x-ray spectroscopy technique as discussed earlier. Whereas, traditional electromagnetic techniques which are involved in measuring CSD and q$_{m}$, are mainly electromagnetic in origin, thereby accounting for the total charge of the ion in the detectors placed at the focal plane, a few meters away from the target. Empirical formalisms are tuned on the basis of the measured data taken from the electromagnetic methods and therefore represent the integral role of the bulk and the surface of the foil. This is the reason why there is a long discrepancy occurs between results of empirical formalism and x-ray method in the low and intermediate energy regime, as reported earlier also \cite{Novikov13,imai9}.\\
Whereas, ETACHA formalism \cite{Rozet} takes account of ionization and capture processes theoretically and does not include surface electron capture processes. Therefore ETACHA predictions ought to be represent the measurements right at the ion-solid interaction and may compare well with the experimental results (this work, x-ray method) in the intermediate energy regime.
It is worth to note that as the projectile energy increases probability of capturing electrons from the target surface start decreasing and at much higher energies (GeV) due to very low interaction time, projectile will not capture electrons from the target surface. Thus at higher energies (GANIL energies) x-ray method, ETACHA predictions and empirical formalism or electromagnetic methods will start following same results due to exclusion of surface effect.

\begin{figure*}[ht]

\includegraphics[scale=.23]{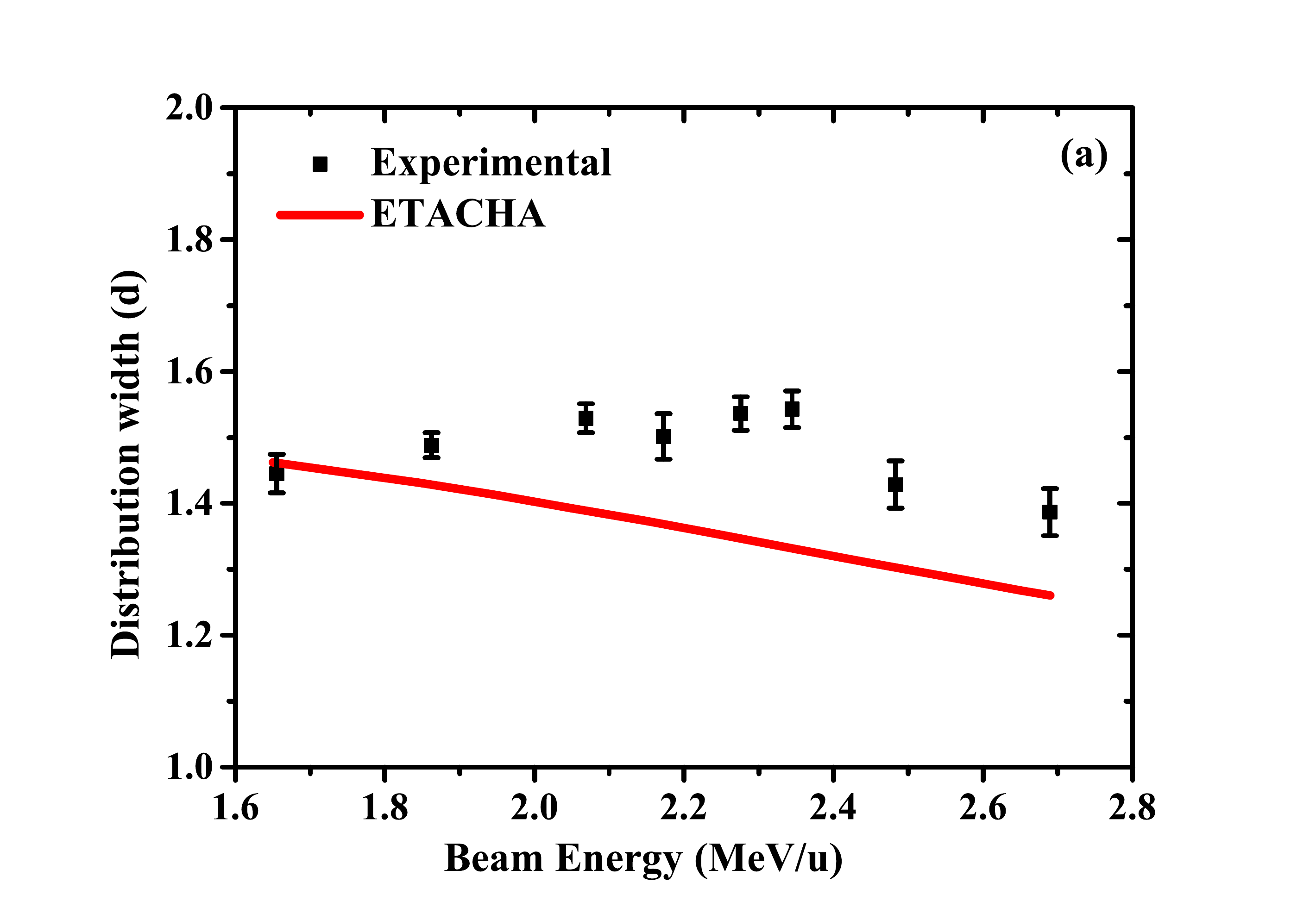}
\includegraphics[scale=.23]{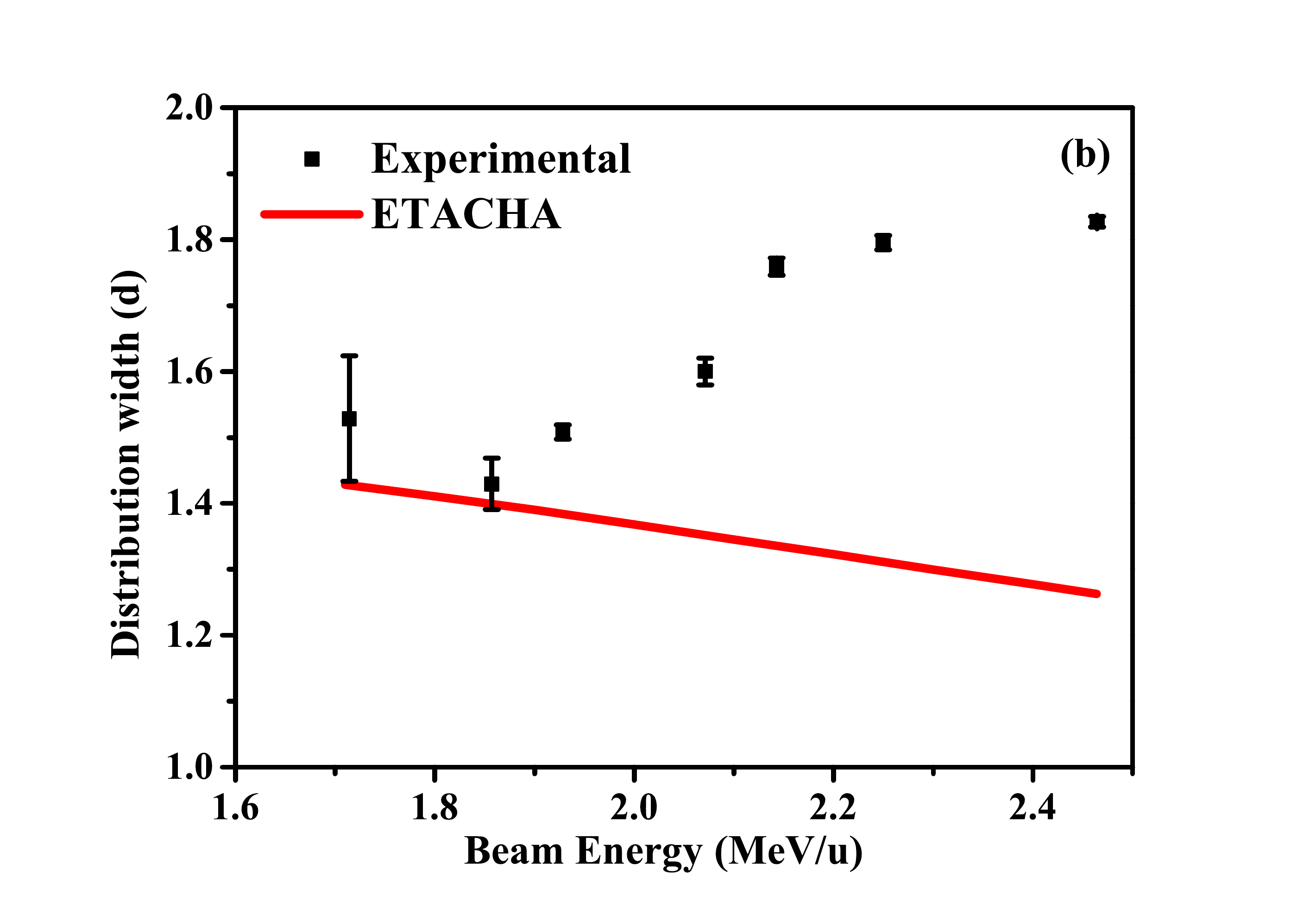}

\caption{\label{fig4} Comparison between experimental and ETACHA distribution widths for (a) $^{58}$Ni on C and (b) $^{56}$Fe on C 
}
\end{figure*}

Next we have evaluated distribution width from the given formula \cite{Novikov14}

\begin{equation}
d^2 = \sum_q (q - q_m)^2 F_q
\end{equation}
and plotted them against beam energies in Fig. \ref{fig4}. Uncertainties in distribution width are determined using standard procedure of propagation of errors. The figure shows good agreement between the measured and ETACHA predicted distribution widths for $^{58}$Ni whereas they depart from each other with higher beam energies for $^{56}$Fe. However, the oscillatory nature in distribution widths is quite common as observed earlier, for example \cite{Novikov13}.
\begin{figure*}[!ht]

\includegraphics[scale=.23]{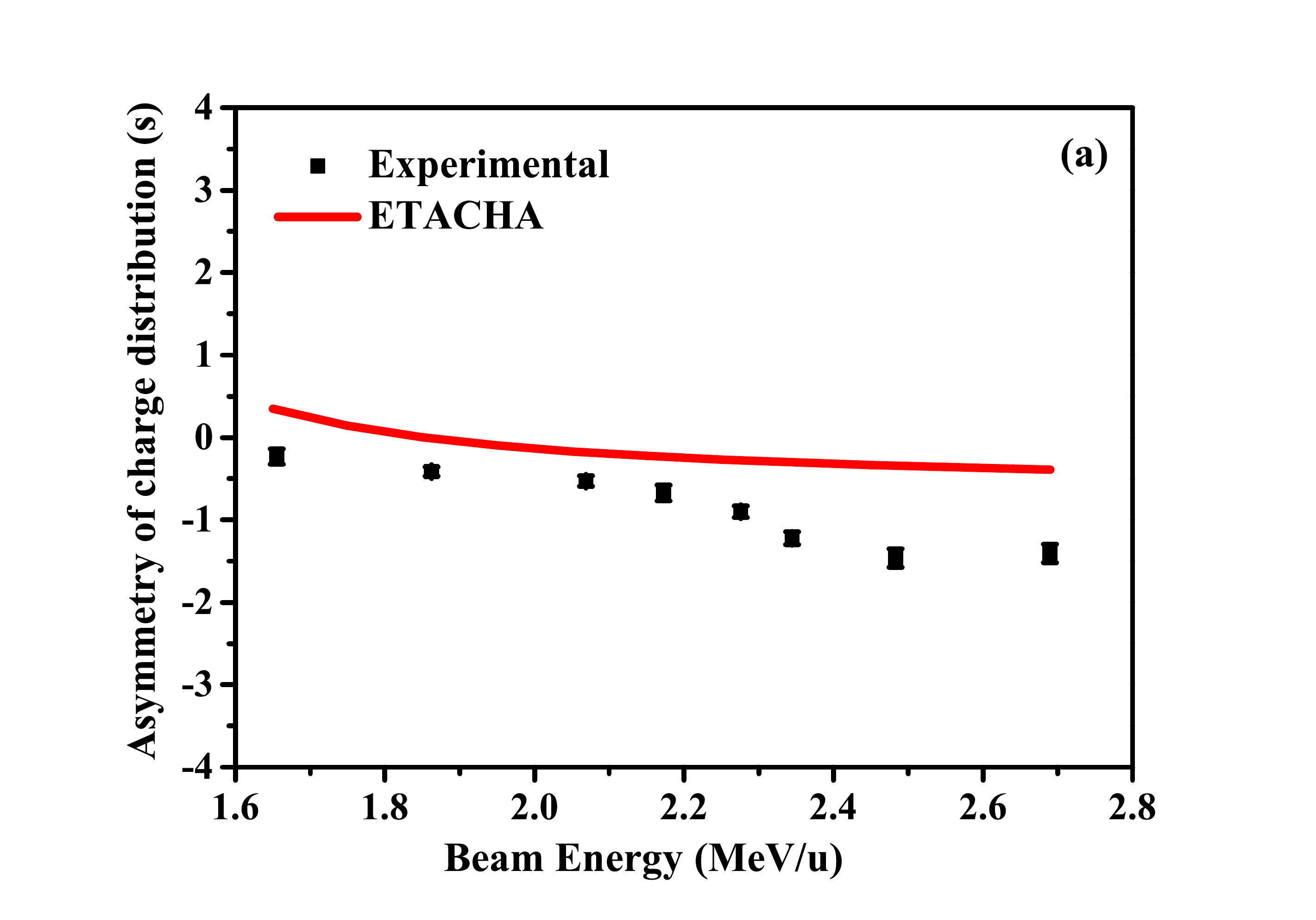}
\includegraphics[scale=.23]{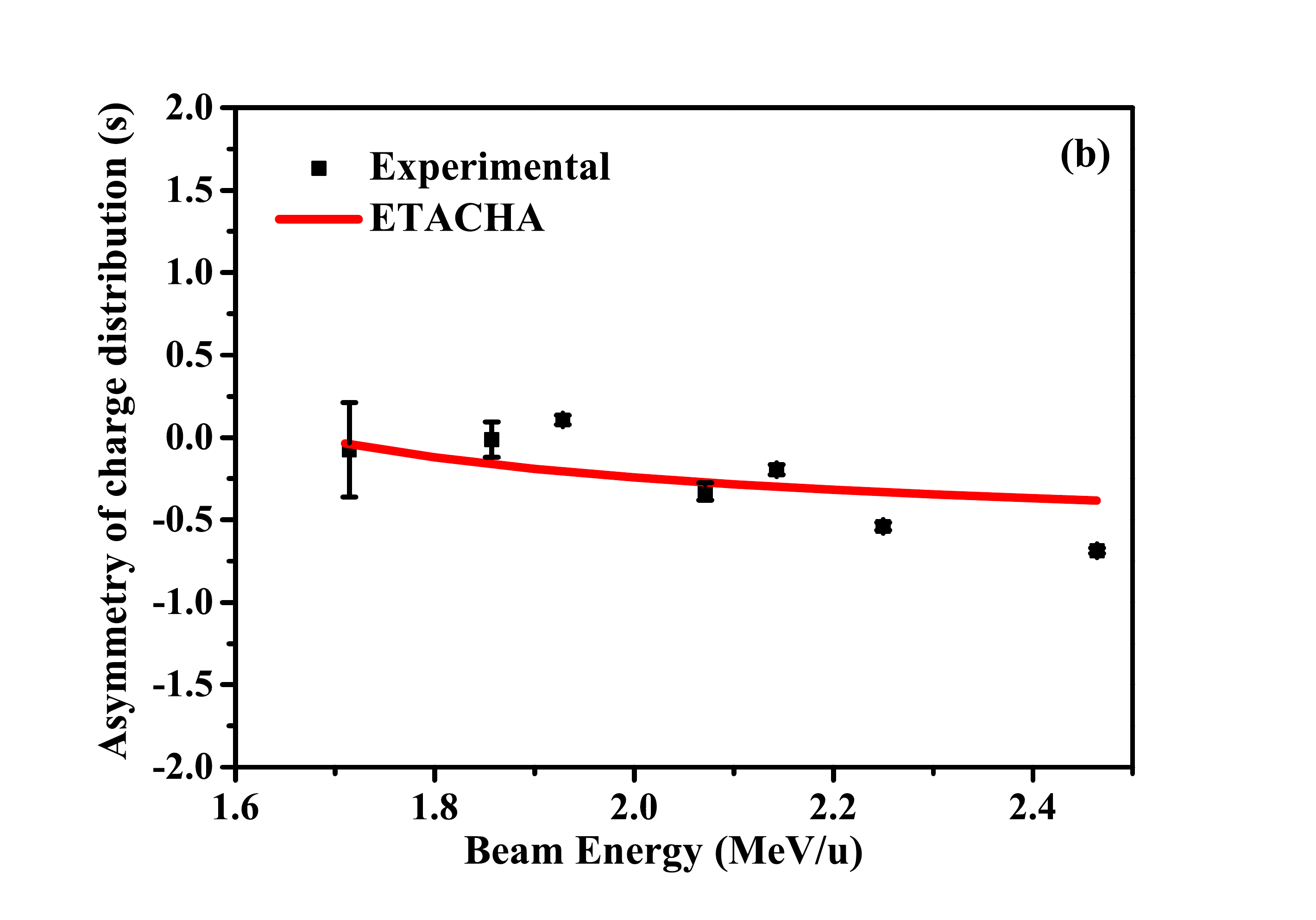}

\caption{\label{fig5} Comparison between experimental and ETACHA asymmetry of charge distribution parameter for (a) $^{58}$Ni on C and (b) $^{56}$Fe on C 
}
\end{figure*}

Further, we have computed the asymmetry of the charge distribution using the formula given by \cite{Novikov14}
\begin{equation}
s = \sum_q (q - q_m)^3 F_q / d^3
\end{equation}
and the data so obtained for experiments as well as ETACHA predictions are plotted in Fig. \ref{fig5}. Again, using standard propagation of errors procedure uncertainties are estimated in the asymmetry of the charge distribution parameters.
 In both the cases of $^{58}$Ni and $^{56}$Fe a decreasing trend with increasing energy is found similar to the ETACHA values.

\section{Conclusion}

In this work we have determined the charge state distribution and its different parameters like mean charge state, distribution width and asymmetry parameter using x-ray spectroscopy technique. This technique is found to be appropriate to segregate the charge state distribution in the bulk from that of the surface by measuring the charge changing phenomena right at the interaction zone. It is shown that ETACHA code \cite{Rozet} represents well the mean charge state measurements starting from energies $\geq$ 2 MeV/u. However, an unusual charge state distribution is observed in the form of Lorentzian distribution in contrast to the prevailing Gaussian distribution as predicted by the ETACHA code. Whereas other charge state distribution parameters are found to showing comparable results to ETACHA predictions \cite{Rozet}. It is observed that the $q_{m}$ data from either this measurement or ETACHA predictions \cite{Rozet} are much higher than that from any empirical formula, which is a clear indication of the multi-electron capture from target surface as reported earlier \cite{Briand1990,Brauning2001,Hay1990,Barat1999}. 

\section*{Acknowledgement}
We would like to acknowledge the co-operation and support received from the Pelletron accelerator staff and all colleagues of Atomic physics group, IUAC, New Delhi. We would also like to gratefully and sincerely thank Dr. M. F Haso\u{g}lu for providing the theoretical fluorescence yield data. PS is thankful to UGC, India for providing the fellowship as financial support to carry out this work.
\section*{References}

\bibliography{mybibfile}

\end{document}